\documentclass[aps,preprint,amssymb]{revtex4}
\usepackage{epsfig}

\def\up{\uparrow}
\def\down{\downarrow}
\def\erfc{{\rm erfc}}

\def\erf{{\rm erf}}

\def\xc{_{\rm xc}}

\def\rv{{\bf r}}

\def\xv{{\bf x}}
\def\Rv{{\bf R}}

\def\beq{\begin{equation}}
\def\eeq{\end{equation}}

\begin{document}
\title{Range separation combined with the Overhauser model: Application to the H$_2$ molecule along the dissociation curve}
\author{Paola Gori-Giorgi$^{1,2}$ and Andreas Savin$^1$}
\affiliation{$^1$ Laboratoire de Chimie Th\'eorique, CNRS UMR76116, Universit\'e Pierre et Marie Curie, 4 Place Jussieu, F-75252 Paris, France\\
$^2$ Afdeling Theoretische Chemie, Vrije Universiteit, De Boelelaan 1083, 1081 HV Amsterdam, The Netherlands}
\date{\today}
\begin{abstract}
	The combination of density-functional theory with other approaches to the many-electron problem through 
	the separation of the electron-electron interaction into a short-range and a long-range contribution (range separation) is a successful strategy, which is raising more and more interest in recent years. We focus here on a range-separated method in which only the short-range correlation energy needs to be approximated, and we model it within the ``extended Overhauser approach".  We consider the paradigmatic case of the H$_2$ molecule along the dissociation curve, finding encouraging results. By means of very accurate variational wavefunctions, we also study how the effective electron-electron interaction appearing in the  Overhauser model should be in order to yield the exact correlation energy for standard Kohn-Sham density functional theory. 
\end{abstract}
\pacs{boh}
\maketitle
\section{Introduction}
\label{intro}
Kohn-Sham (KS) density functional theory (DFT) (see, e.g., \cite{Koh-RMP-99}) is a successful method for electronic structure calculations, thanks to its unique combination of low computational cost and reasonable accuracy. In the Kohn-Sham formalism, the total energy of a many-electron system in the external potential $\hat{V}_{ne}=\sum_i v_{ne}(\rv_i)$ is rewritten as a functional of the one-electron density $\rho(\rv)$,
\beq
E[\rho]=T_s[\rho]+\int d\rv\,v_{ne}(\rv)\,\rho(\rv)+U[\rho]+E\xc[\rho].
\label{eq_Erho}
\eeq
In Eq.~(\ref{eq_Erho}), $T_s[\rho]$ is the kinetic energy of a non-interacting system of fermions (usually called KS system) having the same one-electron density $\rho$ of the physical, interacting, system. The Hartree energy $U[\rho]$ is the classical repulsion energy, $U[\rho]=\frac{1}{2}\int d\rv\int d\rv'\rho(\rv)\rho(\rv')|\rv-\rv'|^{-1}$, and the exchange-correlation functional $E\xc[\rho]$ must be approximated. Minimization of Eq.~(\ref{eq_Erho}) with respect to the spin-orbitals forming the KS determinant lead to the KS equations. Thus, instead of the physical problem, in KS DFT we solve the hamiltonian of a model system of non-interacting fermions, and we recover the energy of the physical system via an approximate functional.

Despite its success in scientific areas ranging from material science to biology, approximate KS DFT is far from being perfect, and many fundamental issues still need to be addressed. In particular, KS DFT encounters difficulties in handling  near-degeneracy correlation effects (rearrangement of electrons within partially filled shells), and in taking into account long-range van der Waals interaction energies (crucial, e.g., for layered materials and biomolecules). In principle, all the shortcomings of KS DFT come from our lack of knowledge of the exchange-correlation functional, and a huge effort is put nowadays in trying to improve the approximations for $E\xc[\rho]$ (for recent reviews see, e.g., \cite{Mat-SCI-02,PerRuzTaoStaScuCso-JCP-05}).  

An alternative strategy to overcome the problems of DFT is range separation: the electron-electron interaction is split into a long-range and a short range part, and the two are treated at different levels of approximation
\cite{StoSav-INC-85,Sav-INC-96,LeiStoWerSav-CPL-97,IikTsuYanHir-JCP-01,KamTsuHir-JCP-02,TawTsuYanYanHir-JCP-04,PolSavLeiSto-JCP-02,TouColSav-PRA-04,AngGerSavTou-PRA-05,TouGorSav-TCA-05,GolWerSto-PCCP-05,GolWerStoLeiGorSav-CP-06,GerAngMarKre-JCP-07,FroTouJen-JCP-07,FroJen-PRA-08,HeyScuErn-JCP-03,HeyScu-JCP-04,HeyScu-JCP-04b,VydHeyKruScu-JCP-06,VydScu-JCP-06,HenIzmScuSav-JCP-07,BaeNeu-PRL-05,BaeLivNeu-CP-06,LivBae-PCCP-07}.
Prof. Hirao has been a pioneer in this field, investigating the effect of range separation on the exchange energy with remarkable success (see, e.g., \cite{IikTsuYanHir-JCP-01,KamTsuHir-JCP-02,TawTsuYanYanHir-JCP-04}). 

The variant of range separation that we consider here \cite{StoSav-INC-85,Sav-INC-96,LeiStoWerSav-CPL-97,PolSavLeiSto-JCP-02,TouColSav-PRA-04,AngGerSavTou-PRA-05,TouGorSav-TCA-05,GolWerSto-PCCP-05,GolWerStoLeiGorSav-CP-06,GerAngMarKre-JCP-07,FroTouJen-JCP-07,FroJen-PRA-08} can be viewed as a way to
remove the constraint that the model system be non-interacting: instead of the KS system, one can define a long-range-only-interacting system (whose wavefunction is thus  multideterminantal) having the same density of the physical system. The remaining part of
the energy is then approximated with a short-range exchange-correlation functional.
The resulting long-range-only hamiltonian, being weakly interacting (and without the electron-electron cusp), can be treated at a reasonable computational cost with standard wavefunction methods: in general, the needed configuration space to achieve good accuracy is small, and often second-order perturbation theory suffices. At the same time, this long-range interaction, albeit small, can make the corresponding wavefunction capture near-degeneracy effects and long-range van der Waals energies. Provided that the energy functionals are correctly redefined, there is no double counting of the energy, and the method is in principle exact, as it is KS DFT. 

As mentioned, this range-separated multideterminant DFT needs an approximation for the short-range exchange-correlation functional. One can follow the same path as for KS DFT: start with the local-spin-density approximation (LSDA), consistently constructed as the difference between the standard LSD functional and the exchange-correlation energy of an electron gas with long-range-only interaction \cite{PazMorGorBac-PRB-06}, and then add gradient corrections (GGA) \cite{GolWerSto-PCCP-05,TouColSav-JCP-05,GolWerStoLeiGorSav-CP-06,FroTouJen-JCP-07}, and eventually
meta-gradient corrections (mGGA). However, this path, which proved highly successful for KS DFT, may not be the best for a scheme in which long-range correlations are explicitly taken into account by wavefunction methods. Indeed, in most cases there is no improvement when passing from LSDA to GGA \cite{GolWerSto-PCCP-05,GolWerStoLeiGorSav-CP-06,GolStoThiSch-PRA-07}, with the exception of hydrogen-bonded complexes \cite{GolLeiManMitWerSto-PCCP-08}.

In recent years we have extended the ``Overhauser model'', an approximate method to calculate the short-range part of the pair density in the uniform electron gas, to systems of nonuniform density \cite{GorSav-PRA-05,GorSav-PM-06,GorSav-IJMPB-07,GorSav-JCTC-07}, finding that it yields an accurate description of the short-range part of the spherically- and system-averaged pair
density (intracule density) of small atoms. In Ref.~\cite{GorSav-IJMPB-07} we have combined the Overhauser equations with the Kohn-Sham equations in a self-consistent way, recovering full CI total energies within 1~mH for the He isoelectronic series. In Ref.~\cite{GorSav-JCTC-07} we have shown that, unlike all the available correlation functionals \cite{KatRoySpr-JCP-06}, the model works equally well for the high-density limit of the He and the Hooke's atom series. 
Thus, on one hand, the Overhauser model seems to be a very good candidate to construct short-range correlation energy functionals. On the other hand, we have tested it only on systems dominated by dynamical correlation: in the He atom, the Overhauser model yields essentially the exact KS correlation energy. However, when we move to systems with strong static correlation we expect the Overhauser model to be unable to yield good results. The combination of the Overhauser model with  range-separated multideterminant DFT seems then natural: it can be viewed as a way to produce an adapted short-range correlation functional for the range-separated multideterminant DFT, or as a way to add the description of static correlation to the Overhauser model.

In this work we combine the Overhauser model with range-separated multideterminant DFT, applying it to the case of the H$_2$ molecule along the dissociation curve, thus analyzing also the case of strong static correlation, as the dissociation limit is approached.  The paper is organized as follows. After briefly reviewing in Secs.~\ref{sec_multDFT} and \ref{sec_Overh} the basic equations of range-separated multideterminant DFT and of the extended Overhauser model, we first analyze in Sec.~\ref{sec_Ovh2exact}, using very accurate variational wavefunctions \cite{RycCenKom-CPL-94,CenKomRyc-CPL-95,CenKut-JCP-96}, how the ``exact'' electron-electron interaction which appears in the Overhauser model (and that it is usually approximated with a physically-motivated interaction) should be as the H$_2$ molecule is stretched. This analysis shows the difficulty of modeling static correlation within the Overhauser model. Since the model is only able to describe correlation, we combine it with a generalized OEP scheme for multideterminant DFT, which is described in Sec.~\ref{sec_mdOEP}. The combination of the two methods is then presented in Sec.~\ref{sec_mdOverh}, with results for the H$_2$ molecule. The last Sec.~\ref{sec_conc} is devoted to conclusions and perspectives.

\section{Multideterminant DFT via range separation}
\label{sec_multDFT}
Hohenberg and Kohn \cite{HohKoh-PR-64} introduced a universal functional  of the density $F[\rho]$, which can be written as a constrained minimum search \cite{Lev-PNAS-79},
\beq
F[\rho]=\min_{\Psi\to\rho}\langle\Psi|\hat{T}+\hat{V}_{ee}|\Psi\rangle.
\label{eq_F}
\eeq
In Eq.~(\ref{eq_F}) the expectation of the kinetic energy operator $\hat{T}=-\frac{1}{2}\sum_i\nabla_i^2$ plus the Coulomb electron-electron repulsion operator $\hat{V}_{ee}=\sum_{i>j}|\rv_i-\rv_j|^{-1}$ is minimized over all wavefunctions yielding the density $\rho$. The universality of the functional $F[\rho]$ stems from the fact that $\hat{T}$ and $\hat{V}_{ee}$ are the same for every electronic system of given particle number $N=\int \rho(\rv)d\rv$. Kohn and Sham \cite{KohSha-PR-65} introduced another functional, $T_s[\rho]$ of Eq.~(\ref{eq_Erho}), by replacing $\hat{V}_{ee}$ in Eq.~(\ref{eq_F}) with zero,
\beq
T_s[\rho]=\min_{\Phi\to\rho}\langle\Phi|\hat{T}|\Phi\rangle,
\label{eq_Ts}
\eeq
and used $T_s[\rho]$ for approximating an important part of $F[\rho]$. In Eq.~(\ref{eq_Ts}), and in the rest of this paper, $\Phi$ denotes a non-interacting wavefunction (thus in the majority of cases a single Slater determinant). Similarly, we can introduce a  functional
$F_{\rm LR}^\mu[\rho]$ for a long-range-only interaction $\hat{W}_{\rm LR}^\mu$ (here chosen using the error function, with the real parameter $\mu$ governing the cutoff of the short-range part),
\beq
\hat{W}_{\rm LR}^\mu=\frac{1}{2}\sum_{i\neq j} \frac{\erf(\mu|\rv_i-\rv_j|)}{|\rv_i-\rv_j|},
\eeq 
by defining
\beq
F_{\rm LR}^\mu[\rho]=\min_{\Psi^\mu\to\rho}\langle\Psi^\mu|\hat{T}+\hat{W}_{\rm LR}^\mu|\Psi^\mu\rangle.
\eeq
In this way we have
\begin{eqnarray}
	\lim_{\mu\to\infty} F_{\rm LR}^\mu[\rho] & = & F[\rho] \\ 
	\lim_{\mu\to 0} F_{\rm LR}^\mu[\rho] & = & T_s[\rho].
\end{eqnarray}
We can then write the total energy of a given many-electron system as
\beq
E[\rho]=F_{\rm LR}^\mu[\rho]+\int d\rv\,v_{ne}(\rv)\,\rho(\rv)+\int d\rv\int d\rv'\rho(\rv)\rho(\rv') \frac{\erfc(\mu|\rv-\rv'|)}{|\rv-\rv'|}+E\xc^\mu[\rho],
\label{eq_Etotmu}
\eeq
where $\erfc(x)=1-\erf(x)$ is the complementary error function. As in KS DFT then, minimization is performed over the wavefunction $\Psi^\mu$,
\begin{eqnarray}
E_0 & = & \min_{\Psi^\mu}\Biggl\{\langle\Psi^\mu|\hat{T}+\hat{W}_{\rm LR}^\mu|\Psi^\mu\rangle+
\int d\rv\,v_{ne}(\rv)\,\rho_{\Psi^\mu}(\rv)+ \nonumber \\ 
& + & \int d\rv\int d\rv'\rho_{\Psi^\mu}(\rv)\rho_{\Psi^\mu}(\rv') \frac{\erfc(\mu|\rv-\rv'|)}{|\rv-\rv'|}+E\xc^\mu[\rho_{\Psi^\mu}]\Biggr\},
\label{eq_E0DFTmd}
\end{eqnarray}
where $\rho_{\Psi^\mu}$ is the density corresponding to $\Psi^\mu$. Eq.~(\ref{eq_E0DFTmd}) yields an effective,
long-range-only-interacting hamiltonian to be solved with a chosen wavefunction method. The short-range
exchange-correlation functional $E\xc^\mu[\rho]$ is then defined as the energy needed to make Eq.~(\ref{eq_Etotmu})
exact, 
\beq
E\xc^\mu[\rho]=F[\rho]-F_{\rm LR}^\mu[\rho]-\int d\rv\int d\rv'\rho(\rv)\rho(\rv') \frac{\erfc(\mu|\rv-\rv'|)}{|\rv-\rv'|}.
\eeq
For instance, the correct LSD approximation to  $E\xc^\mu[\rho]$ is
\beq
E\xc^{\mu, {\rm LSD}}[\rho]=\int\rho(\rv)\left\{\epsilon_{xc}(\rho_\up(\rv),\rho_\down(\rv))-\epsilon^\mu_{xc}(\rho_\up(\rv),\rho_\down(\rv))\right\},
\eeq
where $\epsilon_{xc}(\rho_\up(\rv),\rho_\down(\rv))$ is the exchange-correlation energy per electron of the standard uniform electron gas (with Coulomb electron-electron interaction) and $\epsilon^\mu_{xc}(\rho_\up(\rv),\rho_\down(\rv))$ is the exchange-correlation energy per electron of a uniform electron gas with interaction $\erf(\mu r_{12})/r_{12}$ \cite{PazMorGorBac-PRB-06}.
 
An exact expression for $E\xc^\mu[\rho]$ is found from the adiabatic connection formula \cite{Sav-INC-96,Yan-JCP-98}:
\beq
E\xc^\mu[\rho]=\int_\mu^\infty d \mu' \int_0^\infty 4\pi r_{12}^2 f^{\mu'}(r_{12}) \frac{2}{\sqrt{\pi}}e^{-\mu'^2 r_{12}^2}d r_{12}-\int d\rv\int d\rv'\rho(\rv)\rho(\rv') \frac{\erfc(\mu|\rv-\rv'|)}{|\rv-\rv'|},
\label{eq_adia}
\eeq
where $f^{\mu}(r_{12})$ is the spherically and system-averaged pair density (intracule density) 
obtained by integrating $|\Psi^\mu|^2$ 
over all variables but $r_{12}=|\rv_2-\rv_1|$,
\beq
f^\mu(r_{12}) = \frac{N(N-1)}{2}\sum_{\sigma_1...\sigma_N}
 \int |\Psi^\mu(\rv_{12},\Rv,\rv_3,...,\rv_N)|^2
\frac{d\Omega_{\rv_{12}}}{4\pi} d\Rv d\rv_3...d\rv_N,
\label{eq_intra}
\eeq
with $\Rv=(\rv_1+\rv_2)/2$. The gaussian damping appearing in Eq.~(\ref{eq_adia}) comes from the derivative of the long-range interaction $\erf(\mu r_{12})/r_{12}$ with respect to $\mu$, and shows that the exchange-correlation energy is determined by the short-range part of the intracule density. Notice that when $\mu=0$ Eqs.~(\ref{eq_adia}) yields the KS exchange-correlation energy functional from a nonlinear adiabatic connection \cite{Sav-INC-96,Yan-JCP-98}.

\section{The extended Overhauser model}
\label{sec_Overh}
The extended Overhuaser model consists in writing an effective Schr\"odinger-like equation for the intracule density $f(r_{12})$ of a given system. The basic idea is the following \cite{GorSav-PRA-05,GorSav-PM-06,GorSav-IJMPB-07}. We start with the observation that the intracule density $f(r_{12})$ couples to any electron-electron interaction operator depending only on the interelectronic distance, $\hat{W}=\sum_{i> j}w(|\rv_i-\rv_j|)$,  in the same way as the density $\rho(\rv)$ couples to any local one-body potential operator $\hat{V}=\sum_i v(\rv_i)$, i.e.,
\begin{eqnarray}
\langle\Psi|\hat{W}|\Psi\rangle & = & \int d\rv_{12} f(r_{12}) w(r_{12}), \\
\langle\Psi|\hat{V}|\Psi\rangle & = & \int d\rv \rho(\rv) v(\rv).
\end{eqnarray}
We can then follow the Hohenberg and Kohn  philosophy but with the roles of $\rho(\rv)$ and $f(r_{12})$, and of $\hat{V}_{ne}$ and $\hat{V}_{ee}$, interchanged. That is, in analogy with Eq.~(\ref{eq_F}) we can define a system-dependent functional $G[f]$,
\beq
G[f]=\min_{\Psi\to f}\langle \Psi|\hat{T}+\hat{V}_{ne}|\Psi\rangle,
\label{eq_G}
\eeq
so that the total energy of a given physical system is equal to
\beq
E[f]=G[f]+\int d\rv_{12} \frac{f(r_{12})}{r_{12}}.
\label{eq_Ef}
\eeq
Like Kohn and Sham, we can define another functional by setting $\hat{V}_{ne}$ equal to zero in Eq.~(\ref{eq_G}),
\beq
T_{\rm f}[f]=\min_{\Psi\to f}\langle \Psi|\hat{T}|\Psi\rangle.
\eeq
The functional $T_{\rm f}[f]$ corresponds to the internal kinetic energy of a free (zero external potential) cluster of fermions having the same intracule density of the physical system. The fermions of this cluster interact with an effective interaction $w_{\rm eff}(r_{12})$ which has the same role of the KS potential for the KS system. In practice, this effective interaction must be approximated. Moreover, for $N>2$ electrons the cluster equation become a complicated many-body problem, so that other approximations are needed. As in the original Overhauser model for the uniform electron gas \cite{Ove-CJP-95,GorPer-PRB-01}, we can approximate the cluster equation with a set of radial geminals $g_i(r_{12})$,
\begin{eqnarray}
& & \left[-\frac{1}{r_{12}}\frac{d^2}{dr_{12}^2}r_{12}
+\frac{\ell (\ell+1)}{r_{12}^2}+w_{\rm eff}(r_{12})\right] g_i(r_{12})  =  \epsilon_i\,
g_i(r_{12}) 
\nonumber \\ 
& & \sum_i \vartheta_i|g_i(r_{12})|^2  =  f(r_{12}),
\label{eq_eff}
\end{eqnarray}
whose occupancies $\vartheta_i$ must be defined (e.g., in a determinantal-like way as in the original Overhauser model \cite{GorPer-PRB-01}).
In practice, trying to solve the whole many-electron Schr\"odinger equation by means of Eqs.~(\ref{eq_Ef})-(\ref{eq_eff}) is a daunting task. The idea is rather \cite{GorSav-PRA-05,GorSav-PM-06,GorSav-IJMPB-07} to couple this ``average-pair-density-functional theory'' with a density functional scheme: Eqs.~(\ref{eq_eff}) can be generalized to any $f^\mu(r_{12})$ along the adiabatic connection of DFT. In Refs.~\cite{GorSav-PRA-05,GorSav-IJMPB-07} we started from the effective interaction
$w_{\rm eff}^{\rm KS}(r_{12})$ which, when inserted in Eqs.~(\ref{eq_eff}), gives the intracule density corresponding to the Kohn-Sham system, $f_{\rm KS}(r_{12})$ (that can be obtained from the KS determinant).  We then wrote 
an approximation for $w_{\rm eff}^\mu(r_{12})$ along the long-range adiabatic connection of DFT as
\beq
w_{\rm eff}^\mu(r_{12})=w_{\rm eff}^{{\rm KS}}(r_{12})+w_{\rm eff}^{c,\mu}(r_{12}).
\label{eq_weff}
\eeq
The only term that needs to be approximated is then $w_{\rm eff}^{c,\mu}(r_{12})$, an effective interaction that should essentially ``tell'' to the intracule density that, while the electron-electron interaction is turned on (i.e. as $\mu$ increases), the one-electron density $\rho(\rv)$ does not change. As the information on $\rho(\rv)$ has been ``washed away'' in the integration over the center of mass $\Rv$ of Eq.~(\ref{eq_intra}), this constraint can be imposed only in an approximate way. For two-electron atoms, for which Eq.~(\ref{eq_eff}) is exact with one geminal \cite{GorSav-PM-06}, $g=\sqrt{f}$, a simple approximation for  $w_{\rm eff}^{c,\mu}(r_{12})$ is \cite{GorSav-PRA-05,GorSav-IJMPB-07,GorSav-JCTC-07}
\beq
w_{\rm eff}^{c,\mu}(r_{12})=\frac{\erf(\mu\, r_{12})}{r_{12}}-\left(\frac{4\pi}{3}\overline{r}_s^3\right)^{-1}\int_{|\xv|\leq \overline{r}_s}\frac{\erf(\mu |\rv_{12}-\xv|)}{|\rv_{12}-\xv|}\,d\xv,
\label{eq_wc}
\eeq
where $\overline{r}_s$ is a screening length associated to the radius of a sphere containing on average one electron \cite{Ove-CJP-95,GorPer-PRB-01,GorSeiSav-PCCP-08}. The physical idea behind Eq.~(\ref{eq_wc}) is to mimic  the constraint of fixed one-electron density by screening the electron-electron interaction over a length associated to the ``space'' available to each electron (which is determined by the density).
Indeed, for the He isoelectronic series Eqs.~(\ref{eq_adia}), (\ref{eq_eff}) and (\ref{eq_wc}), combined self-consistently with the Kohn-Sham equations, recover the full CI total energy within 1 mH \cite{GorSav-PRA-05,GorSav-IJMPB-07,GorSav-JCTC-07}.
\begin{figure}
\includegraphics[width=8cm]{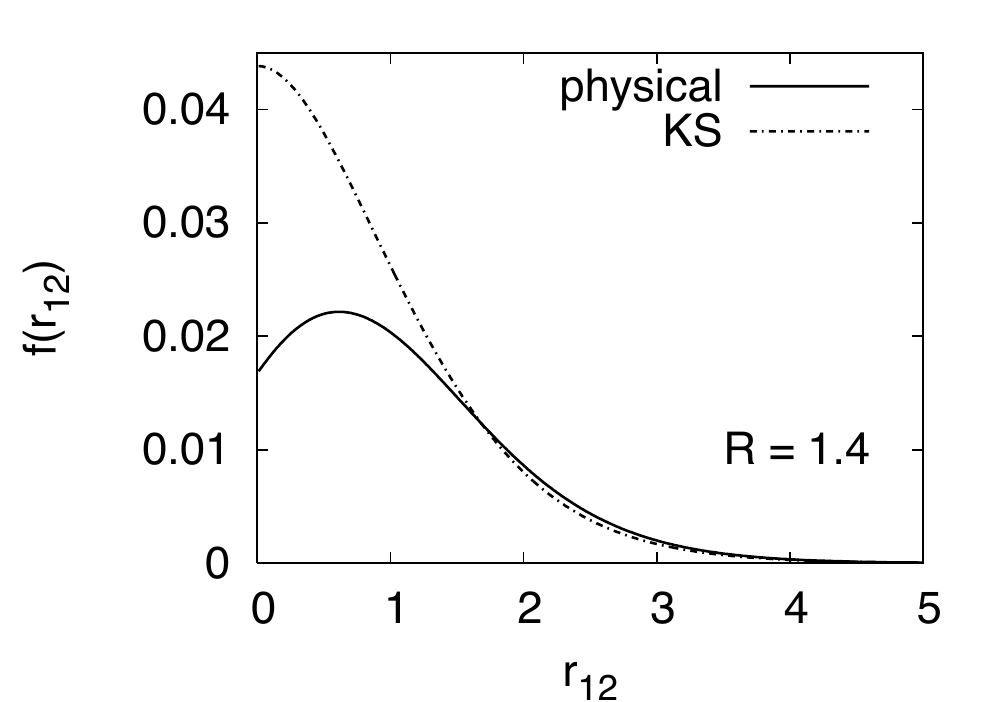}
\includegraphics[width=8cm]{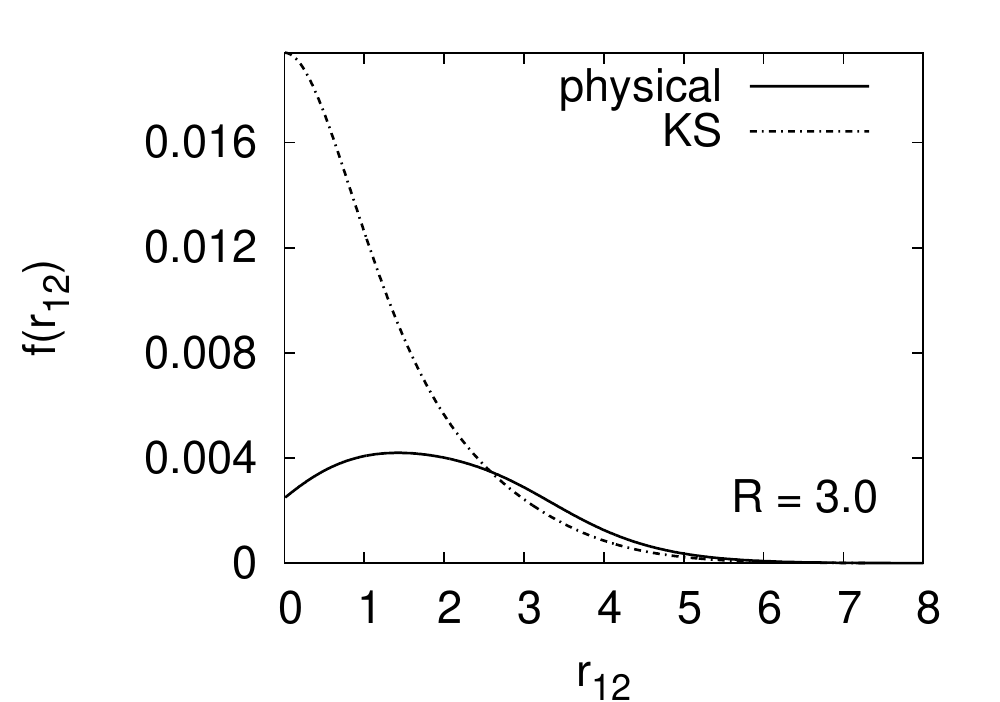}
\includegraphics[width=8cm]{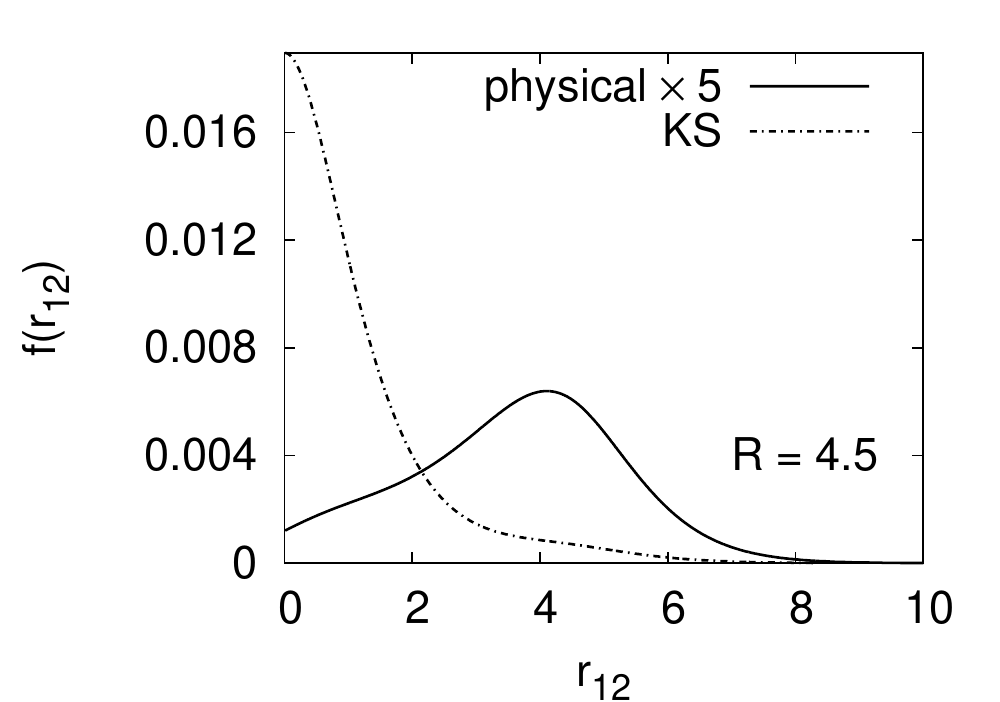}
\includegraphics[width=8cm]{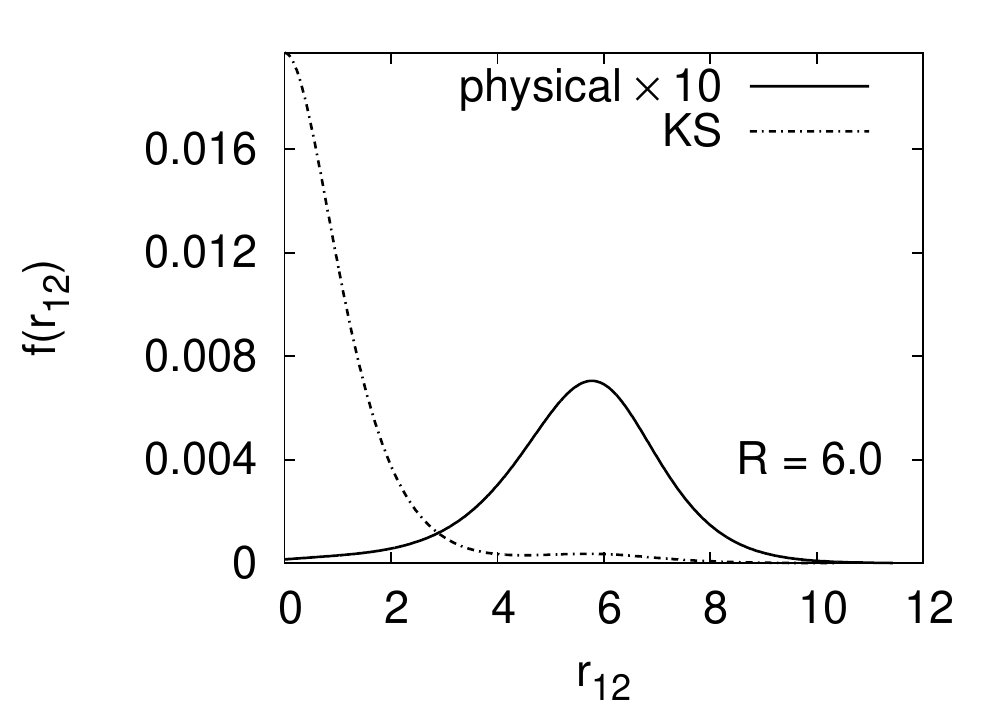} 
\caption{Intracule densities $f(r_{12})$ for the H$_2$ molecule at different internuclear distances $R$ for the physical system (from the accurate variational wavefunction described in Subsec.~\ref{subsec_cencek}) and for the KS system (from the density corresponding to the same accurate variational wavefunctions).}
\label{fig_f}
\end{figure}
\section{The Overhauser model for the H$_2$ molecule: how things should be}
\label{sec_Ovh2exact}
For a closed-shell physical electronic system (atom, molecule) with $N=2$ particles, the Schr\"odinger equation describing the internal degrees of freedom of a cluster of fermions having the same intracule density $f(r_{12})$ is exactly given by \cite{GorSav-PM-06,GorSav-IJMPB-07,GorSav-JCTC-07}
\beq
\left[-\frac{1}{r_{12}}\frac{d^2}{d r_{12}^2} r_{12}+w_{\rm eff}(r_{12})\right]\sqrt{f(r_{12})}=\epsilon \sqrt{f(r_{12})}.
\label{eq_eff2ele}
\eeq
As a first study, we calculate and analyze the ``exact'' Overhauser interaction $w_{\rm eff}(r_{12})$ at full coupling strength (i.e., for electron-electron interaction $1/r_{12}$, corresponding to $\mu=\infty$) for the H$_2$ molecule at different values of the internuclear distance $R$, and we compare it with the approximation of Eq.~(\ref{eq_wc}). To this purpose, we need extremely accurate
intracule densities $f(r_{12})$, which are described in the next Subsec.~\ref{subsec_cencek}.

\subsection{Intracule densities from accurate variational wavefunctions}
\label{subsec_cencek}
We use the accurate variational wavefunctions of Refs.~\cite{RycCenKom-CPL-94,CenKomRyc-CPL-95,CenKut-JCP-96}, which are expanded in explicitly correlated gaussian geminals,
\begin{eqnarray}
	\Psi(\rv_1,\rv_2) & = & (1+\hat{P}_{12})(1+\hat{i}_e)\sum_{k=1}^K c_k\psi_k(\rv_1,\rv_2) 
	\label{eq_expa}\\
	\psi_k(\rv_1,\rv_2) & = & e^{-\alpha_k|\rv_1-\rv_{Ak}|^2}e^{-\beta_k|\rv_2-\rv_{Bk}|^2}e^{-\gamma_k r_{12}^2},
\label{eq_gaussgem}
\end{eqnarray}
where $\rv_{Ak}$ and $\rv_{Bk}$ are centers that lie on the internuclear axis, $\hat{P}_{12}$ means permutation of $\rv_1$ and $\rv_2$, and $\hat{i}_e$ is the inversion operator with respect to the center of the molecule. The parameters appearing in Eqs.~(\ref{eq_expa})-(\ref{eq_gaussgem}) are determined variationally by minimizing the energy with
the conjugate gradient method (for more details on the wavefunction and the algorithms employed, see Refs.~\cite{RycCenKom-CPL-94,CenKomRyc-CPL-95,CenKut-JCP-96}). The expansion length $K=1200$ in Eq.~(\ref{eq_expa})
is used, resulting in energies with the extraordinary accuracy of $10^{-10}$ Hartree.

\begin{figure}
\includegraphics[width=8cm]{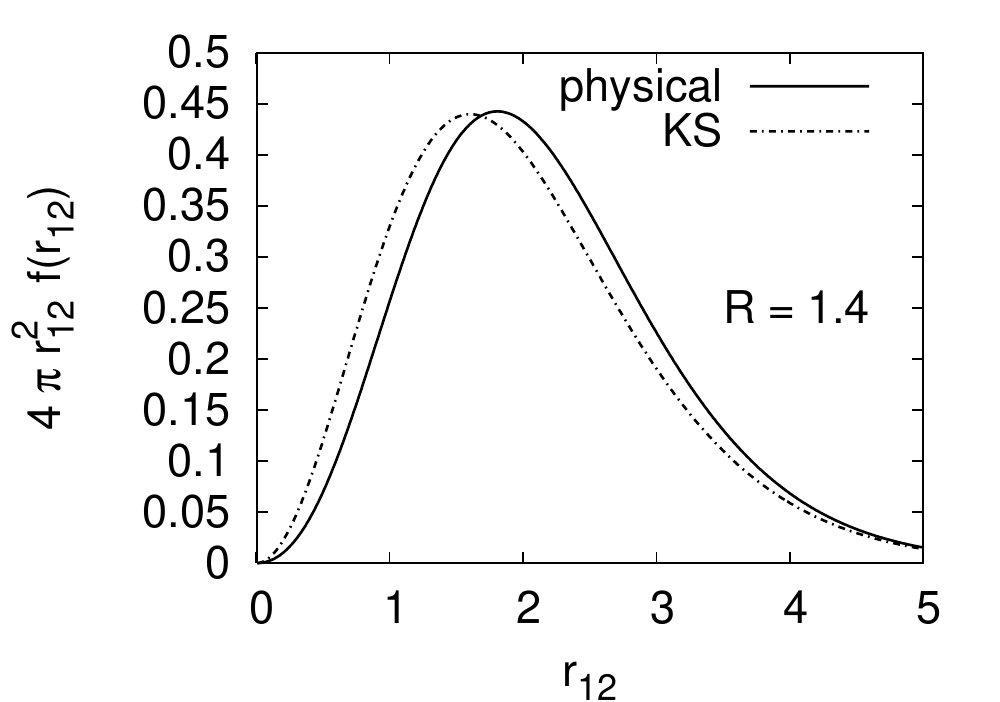}
\includegraphics[width=8cm]{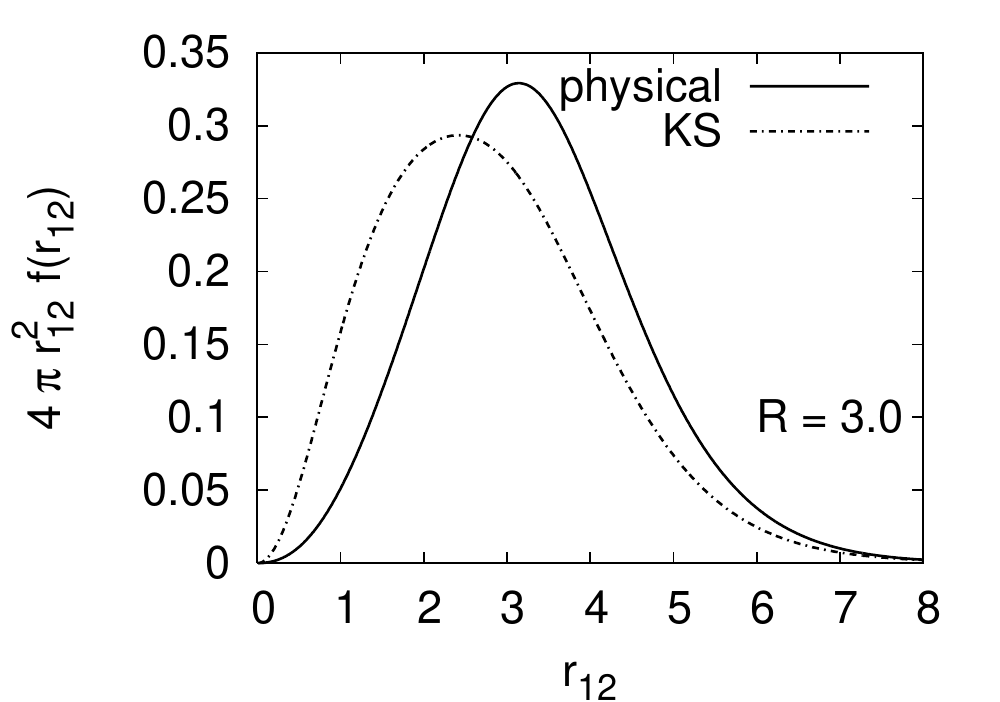}
\includegraphics[width=8cm]{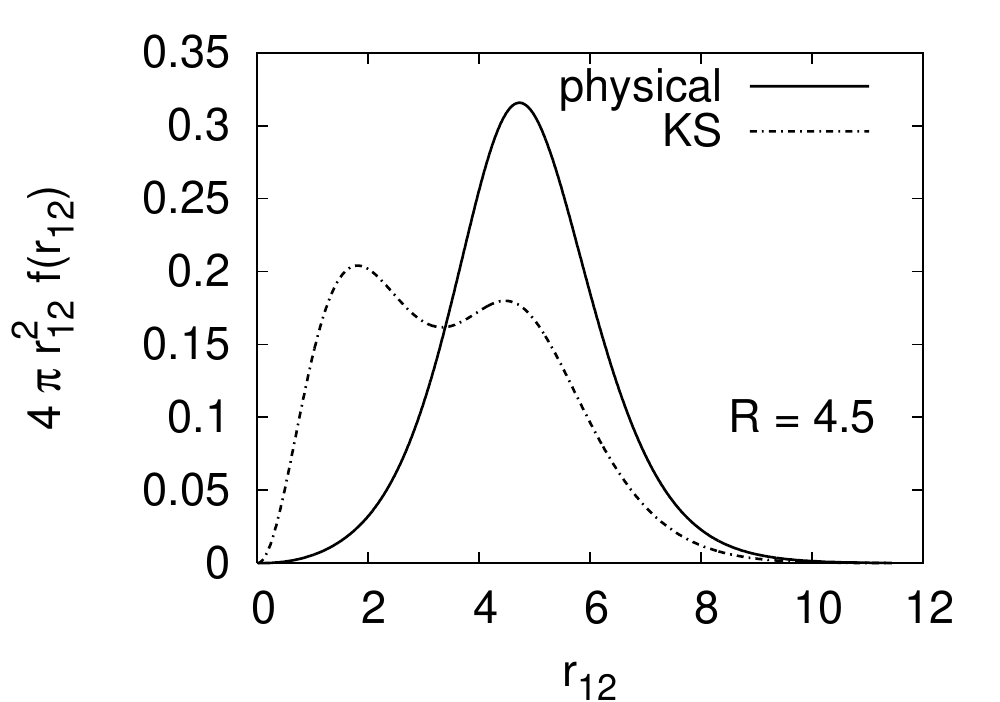}
\includegraphics[width=8cm]{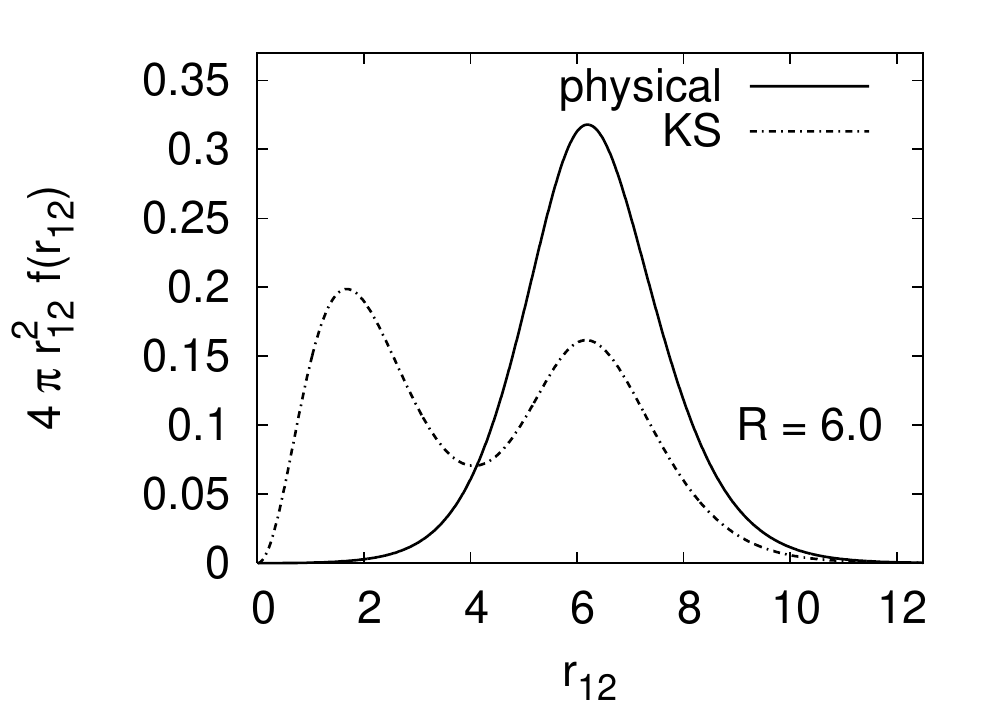} 
\caption{The same intracule densities of Fig.~\ref{fig_f} multiplied by the volume element $4\pi r_{12}^2$.}
\label{fig_fu2}
\end{figure}

\begin{figure}
\includegraphics[width=8.5cm]{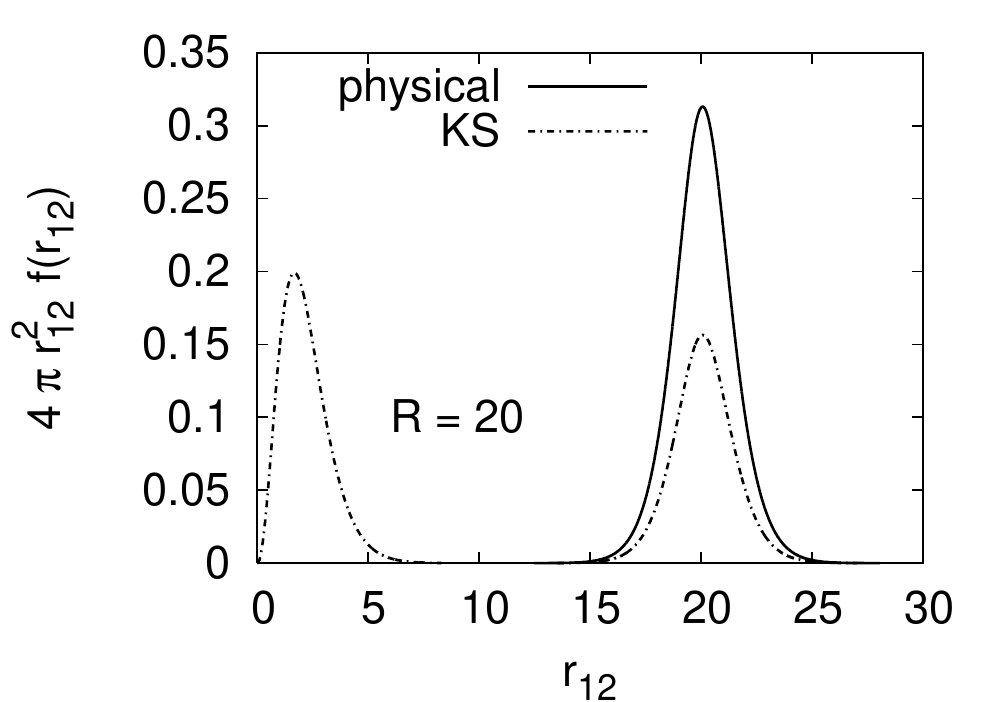}
\caption{The intracule density $f(r_{12})$ multiplied by the volume element $4\pi r_{12}^2$ for the H$_2$ molecule in the extreme stretched case $R=20$. The physical $f(r_{12})$ has been calculated from the simple Heitler-London wavefunction, and $f_{\rm KS}(r_{12})$ from its corresponding density.}

\label{fig_HL}
\end{figure}

The intracule densities $f(r_{12})$ from these extremely accurate wavefunctions can be easily calculated, since all the needed integrals are analytic. We also calculated the one electron densities $\rho(\rv)$, and the intracule densities $f_{\rm KS}(r_{12})$ corresponding to the KS system, which can be obtained by inserting in Eq.~(\ref{eq_intra}) the KS wavefunction $\frac{1}{2}\sqrt{\rho(\rv_1)}\sqrt{\rho(\rv_2)}$. In Fig.~\ref{fig_f} we
show the intracule densities $f(r_{12})$ and $f_{\rm KS}(r_{12})$ for the internuclear distances $R=1.4$, $3.0$, $4.5$ and $6.0$ a.u. Although mathematically the wave 
function of Eqs.~(\ref{eq_expa})-(\ref{eq_gaussgem}) is cuspless, we see that the very elaborate ansatz 
permits to describe 
the exact linear behaviour of the intracule density for $r_{12}\to 0$, up to extremely short 
distances. Fig.~\ref{fig_fu2} shows the same quantities multiplied by the volume element $4\pi r_{12}^2$. This figure better visualizes the transition from dynamical to static correlation. In Fig.~\ref{fig_HL} we also report the same quantities in the extreme stretched case, $R=20$, obtained from the simple Heitler-London wavefunction.
\begin{figure}
\includegraphics[width=8cm]{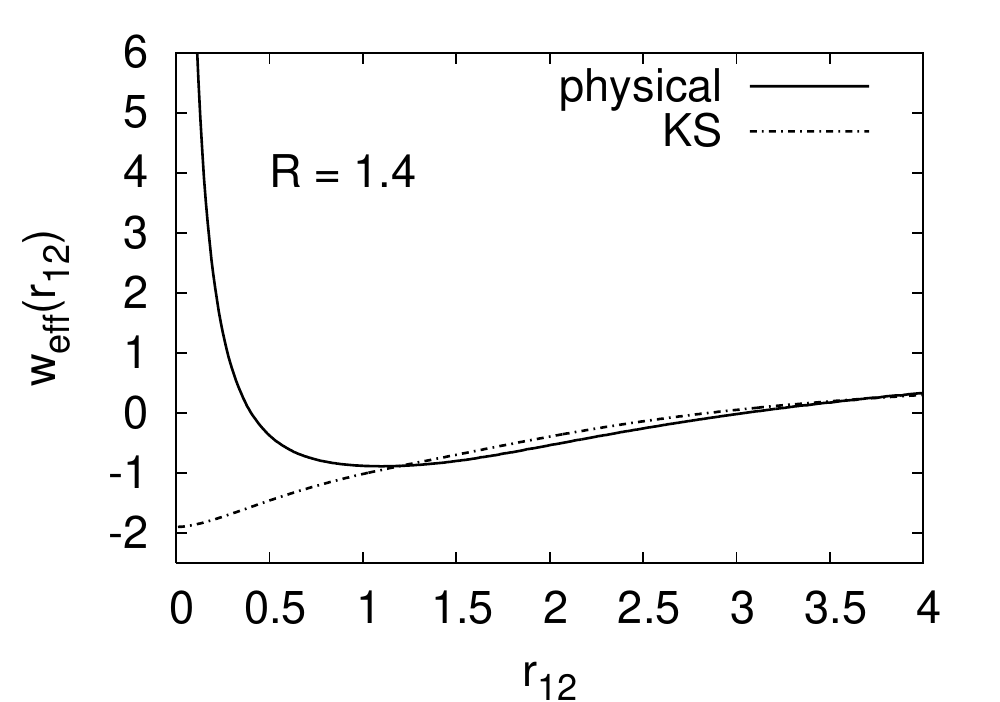}
\includegraphics[width=8cm]{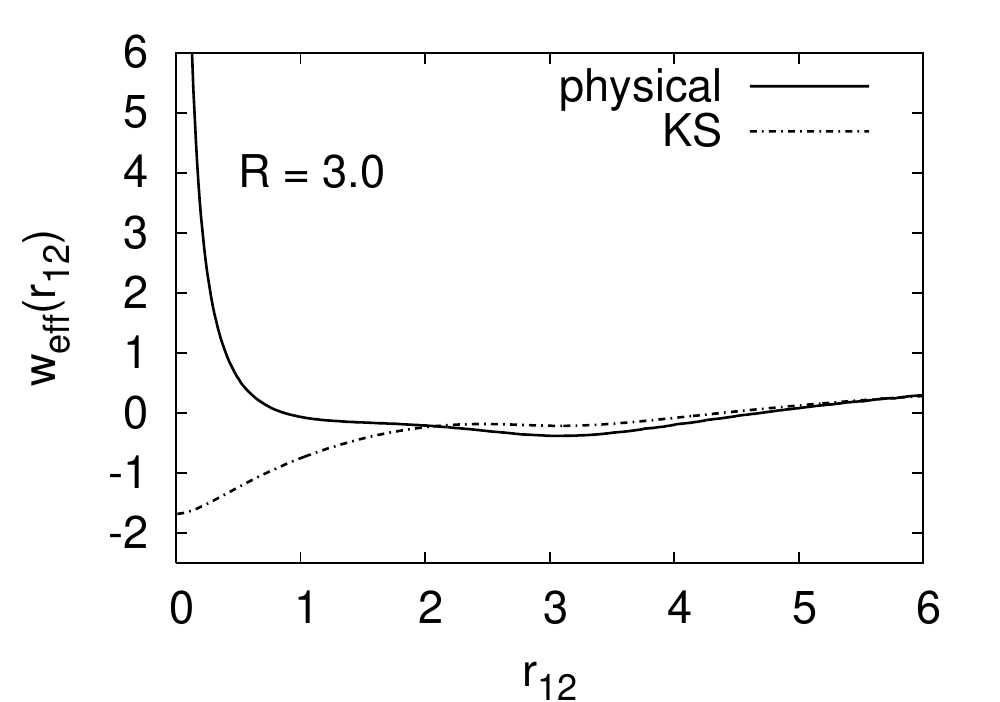}
\includegraphics[width=8cm]{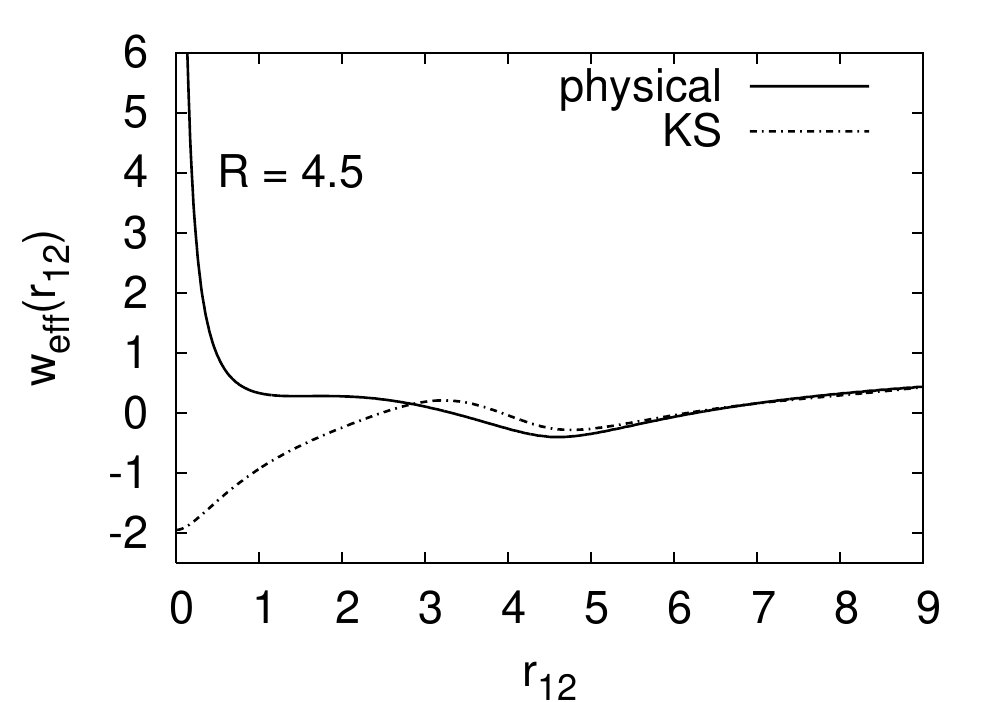}
\includegraphics[width=8cm]{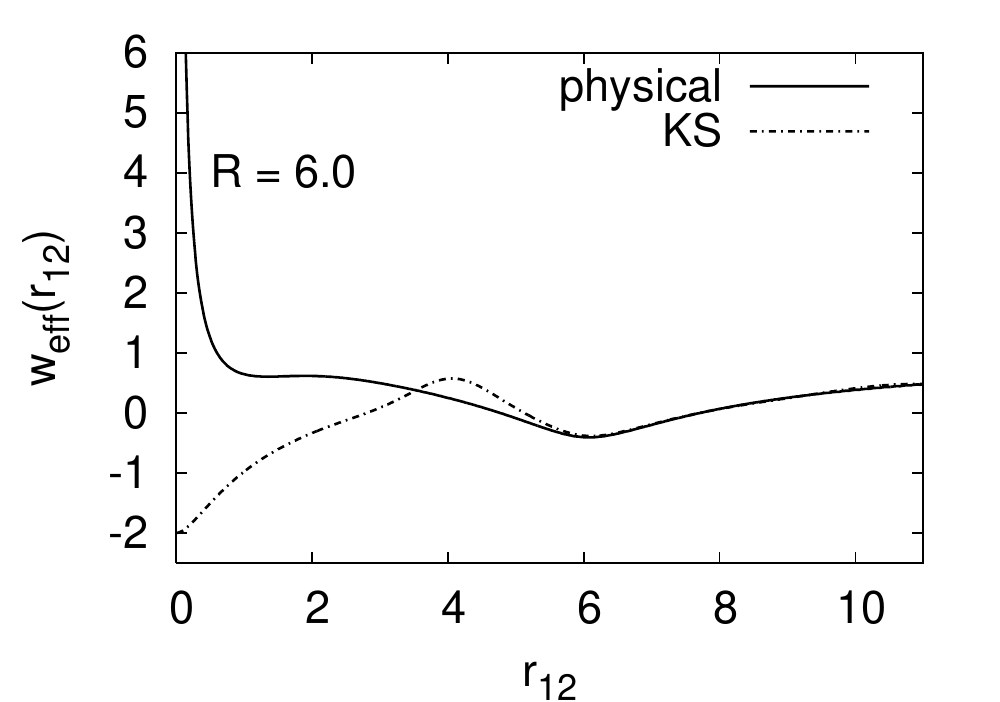} 
\caption{The ``exact'' effective Overhauser interaction $w_{\rm eff}(r_{12})$ for the intracule densities of the physical and of the KS systems of Fig.~\ref{fig_f}.}
\label{fig_weff}
\end{figure}
\begin{figure}
\includegraphics[width=8cm]{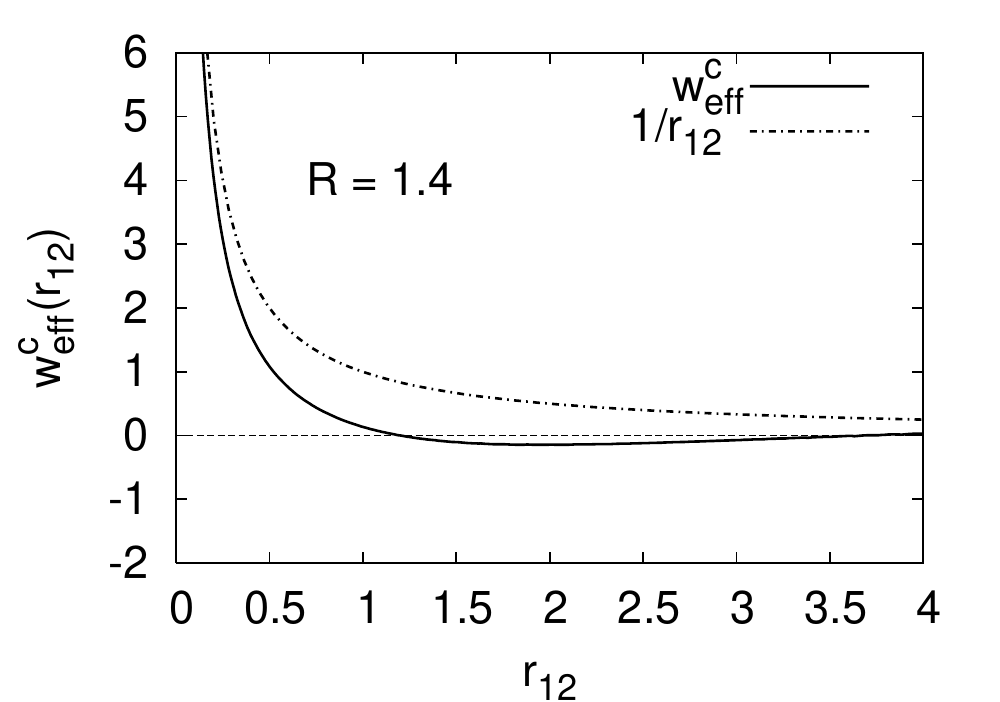}
\includegraphics[width=8cm]{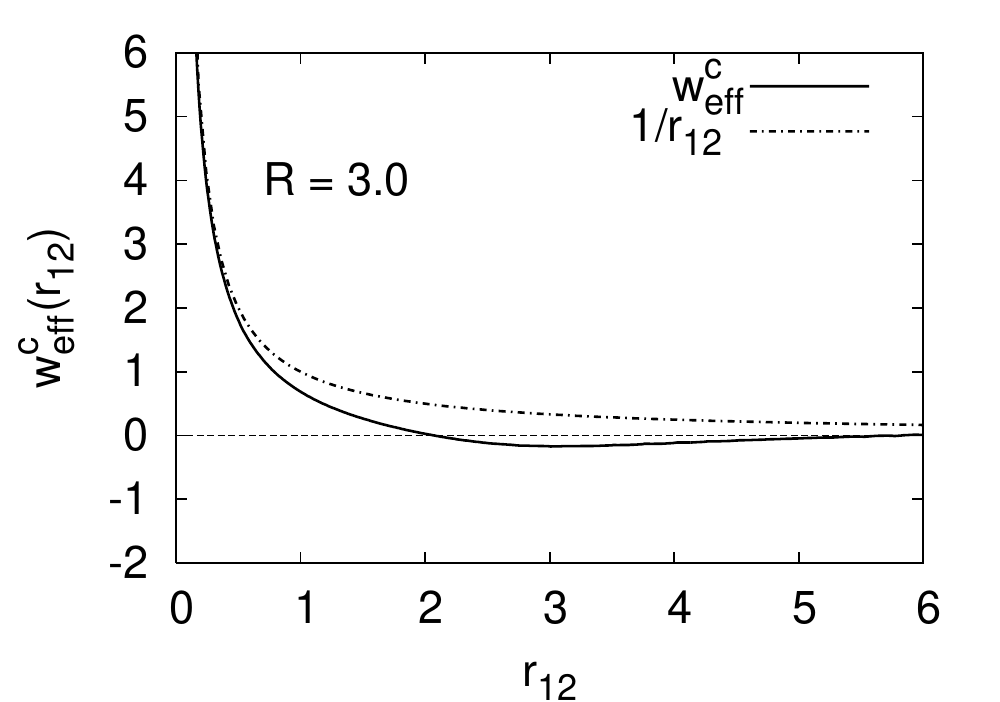}
\includegraphics[width=8cm]{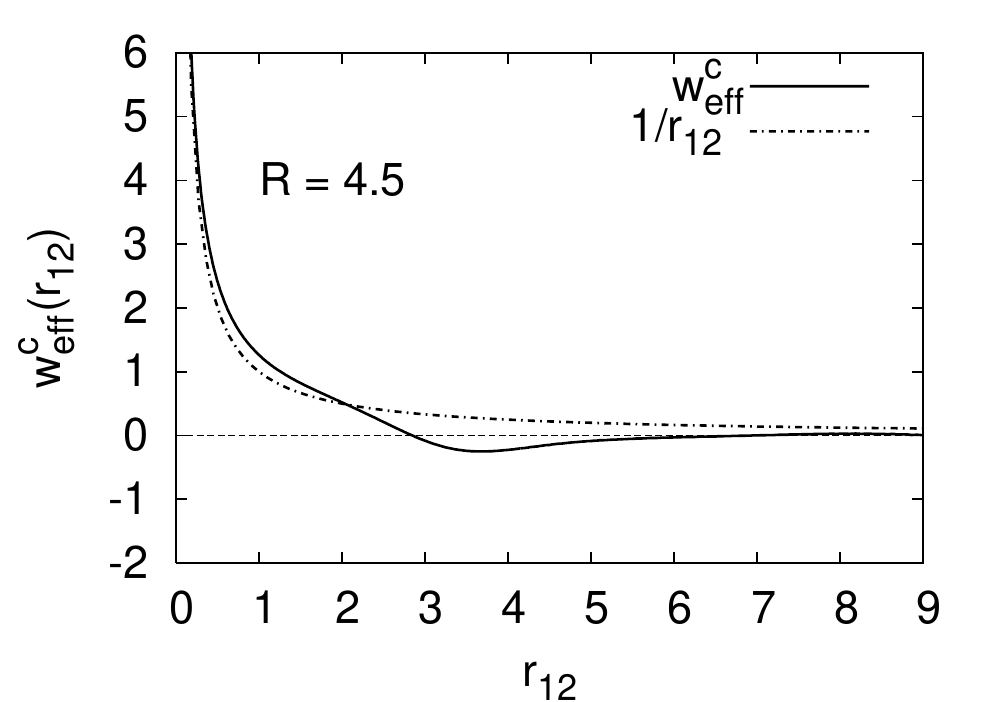}
\includegraphics[width=8cm]{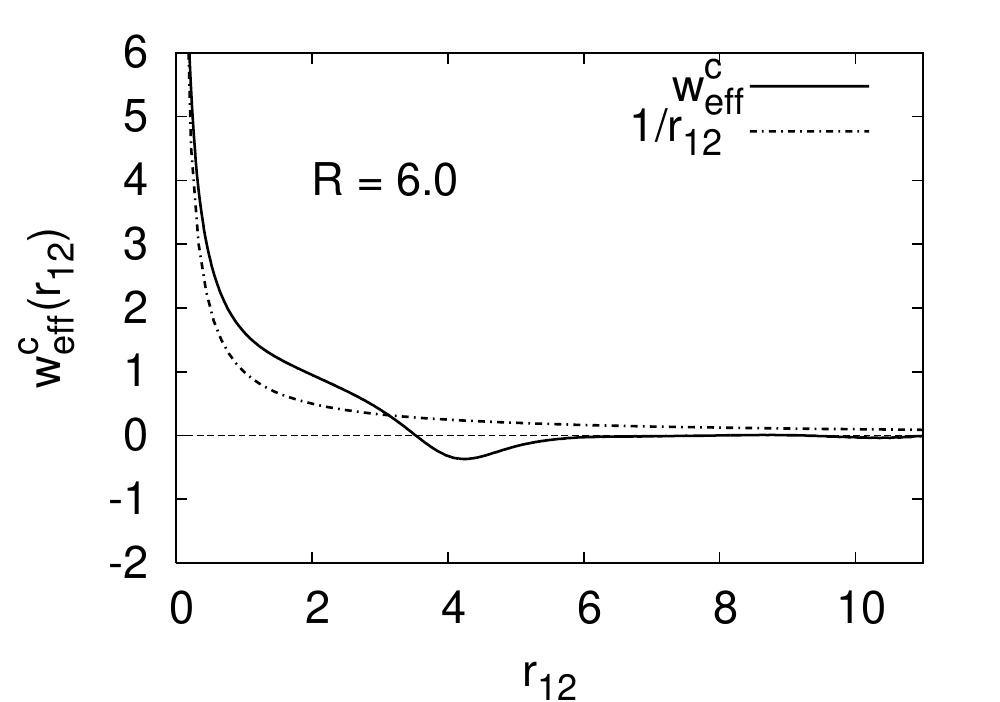} 
\caption{The difference $w_{\rm eff}^c(r_{12})$ between the ``exact'' effective Overhauser interaction for the intracule density of the physical and of the KS systems of Fig.~\ref{fig_weff}. The Coulomb repulsion $1/r_{12}$ is also reported.}
\label{fig_wc}
\end{figure}
\begin{figure}
\includegraphics[width=8cm]{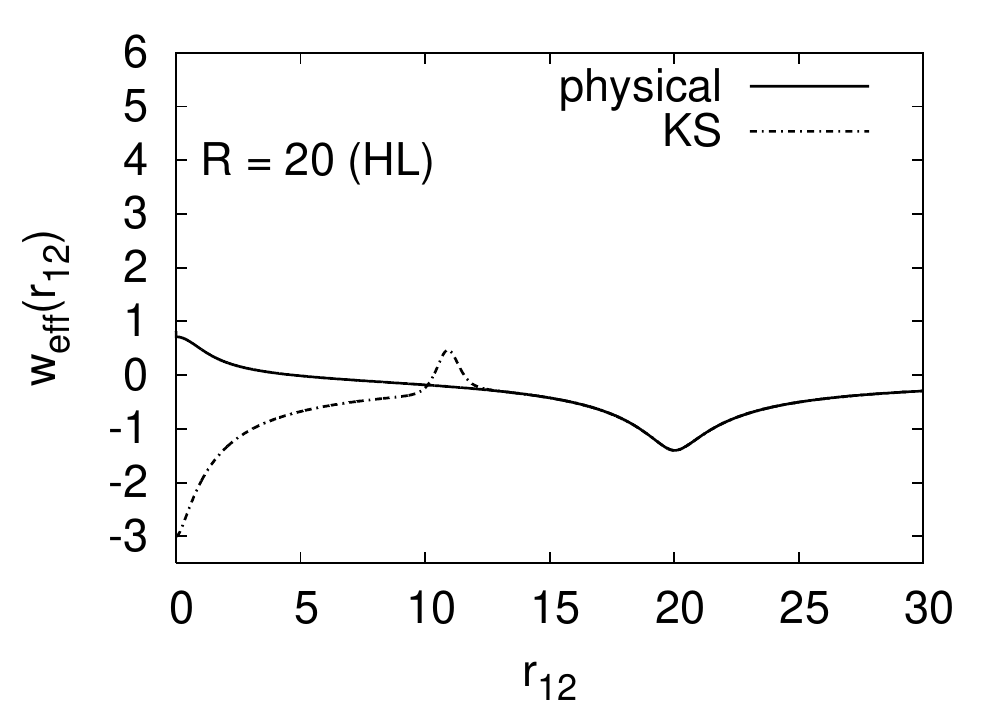}
\includegraphics[width=8cm]{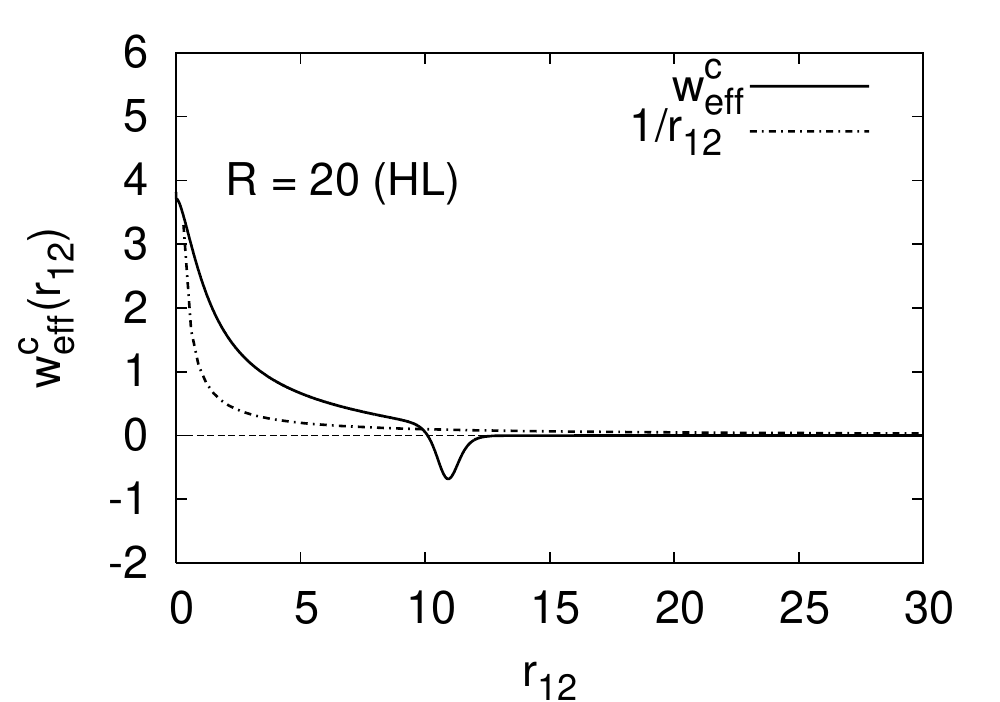}
\caption{Same as Figs.~\ref{fig_weff}-\ref{fig_wc} for the extreme stretched molecule, using the simple Heitler-London wavefunction.}
\label{fig_wHL}
\end{figure}
\subsection{Accurate Overhauser potentials}
From the accurate intracule densities of the previous subsection we can calculate, by inversion, the corresponding ``exact'' Overhauser interaction $w_{\rm eff}(r_{12})$,
\beq
w_{\rm eff}(r_{12})=\frac{1}{\sqrt{f(r_{12})}}\frac{1}{r_{12}}\frac{d^2}{d r_{12}^2}\left(r_{12}\sqrt{f(r_{12})}\right)+{\rm const.}
\label{eq_inv}
\eeq
The inversion of Eq.~(\ref{eq_inv}) is done numerically, by finite differences. In Fig.~\ref{fig_weff} we report the effective Overhauser interactions that, when inserted in Eq.~(\ref{eq_eff2ele}), give the physical and the KS intracule, corresponding, respectively, to $\mu=\infty$ and $\mu=0$ along the long-range adiabatic connection of Sec.~\ref{sec_multDFT} (or to $\lambda=1$ and $\lambda=0$ along the usual linear adiabatic connection in which $\hat{V}_{ee}$ is simply multiplied by $\lambda$). We see that $w_{\rm eff}(r_{12})$ for large $r_{12}$ goes to the same constant for both the KS and the physical system, as it should be \cite{GorSav-IJMPB-07} (of course if we go to too large $r_{12}$ we start to observe the wrong harmonic wall due to the gaussian asymptotic decay of our wavefunction). The difference between the effective Overhauser interaction for the physical and the KS system gives $w_{\rm eff}^{c,\mu\to\infty}\equiv w_{\rm eff}^c$ of Eq.~(\ref{eq_weff}), and is reported in Fig.~\ref{fig_wc}, where also the Coulomb repulsion $1/r_{12}$ is shown. From this figure, we see that, when the system is still dominated by dynamical correlation, as in the case $R=1.4$ and $R=3$, $w_{\rm eff}^c(r_{12})$ is essentially a screened Coulomb interaction. That is, for short-range it behaves as $1/r_{12}$, and then for large $r_{12}$  goes to zero much faster than $1/r_{12}$. In such cases, the approximation of Eq.~(\ref{eq_wc}), which at $\mu=\infty$ reads

\begin{eqnarray}
  w_{\rm eff}^c(r_{12})    =  & \frac{1}{r_{12}}
+\frac{r_{12}^2}{2\overline{r}_s^3}-\frac{3}{2 \overline{r}_s}
  \qquad  &  r_{12}\le \overline{r}_s \nonumber  \\
 w_{\rm eff}^c(r_{12})     =  & 0    & r_{12}>\overline{r}_s. \label{eq_wcCoulomb}
\end{eqnarray}
can work reasonably well, with a screening length $\overline{r}_s\sim R$. However, as $R$ grows and the system starts to be dominated by static correlation, we see that the approximation of Eq.~(\ref{eq_wcCoulomb}) cannot work: the ``exact'' $w_{\rm eff}^c(r_{12})$ still decays much faster than $1/r_{12}$ for large $r_{12}$, but at short range is more repulsive than the Coulomb interaction! I.e., we need an ``overscreened'' interaction. This is completely evident in the extreme stretched case $R=20$ of Fig.~\ref{fig_wHL}, again obtained from the simple Heitler-London wavefunction. 

\section{Generalized optimized effective potential method for multideterminant DFT}
\label{sec_mdOEP}
In recent years,
the focus of a large part of the scientific community working on improving the approximations for $E_{xc}[\rho]$ has shifted from seeking explicit functionals of the density (like the generalized gradient approximations), to implicit functionals, typically using the exact exchange $E_{\rm x}[\rho]$, which is only explicitly known in terms of the Kohn-Sham orbitals $\phi_i(\rv)$. The corresponding Kohn-Sham potential must then be computed with the optimized effective potential (OEP) method (for a recent review, see \cite{KumKro-RMP-08}). The OEP scheme can be generalized to the multideterminant range-separated DFT  by first noticing that we
can divide $E_{xc}^\mu[\rho]$ into exchange and correlation in two different ways \cite{TouGorSav-TCA-05}: we can define the exchange energy with respect to the KS determinant $\Phi$,
\beq
E_x^\mu[n]  =  
\langle \Phi |\hat{V}_{ee}-\hat{W}_{\rm LR}^\mu|\Phi\rangle
-\int d\rv\int d\rv'\rho(\rv)\rho(\rv') \frac{\erfc(\mu|\rv-\rv'|)}{|\rv-\rv'|},
\eeq
and then define the usual correlation energy functional $E^\mu_c[\rho]$ as the energy missed by the KS wavefunction,
\beq
E^\mu_c[\rho]=E^\mu_{\rm xc}[\rho]-E_{\rm x}^\mu[\rho],
\label{eq_Eccomp}
\eeq
but we can also define a multideterminantal (md) exchange 
functional \cite{TouGorSav-TCA-05} by using the wavefunction $\Psi^\mu$,
\beq
E_{\rm x, md}^\mu[\rho]  =  
\langle \Psi^\mu |\hat{V}_{ee}-\hat{W}_{\rm LR}^\mu|\Psi^\mu\rangle-
\int d\rv\int d\rv'\rho(\rv)\rho(\rv') \frac{\erfc(\mu|\rv-\rv'|)}{|\rv-\rv'|},
\label{eq_Exmd}
\eeq
and then a corresponding correlation energy that recovers the energy missed by $\Psi^\mu$ (which is smaller than the energy missed by the KS determinant $\Phi$),
\beq
E^\mu_{\rm c, md}[\rho]=E^\mu_{\rm xc}[\rho]-E_{\rm x, md}^\mu[\rho].
\label{eq_Ecmd}
\eeq
Then, with this latter definition of the correlation energy, the generalized OEP-like scheme for multideterminant DFT becomes \cite{TouGorSav-TCA-05}
\beq
E_0=\inf_{v^\mu}\left\{\langle\Psi^\mu_{v^\mu}| \hat{T}+\hat{V}_{ee}+\hat{V}_{ne}
|\Psi^\mu_{v^\mu}\rangle+E^\mu_{\rm c, md}[\rho_{\Psi^\mu_{v^\mu}}]\right\},
\label{eq_OEPmd}
\eeq
where $\Psi^\mu_{v^\mu}$ is obtained by solving the Schr\"odinger equation
corresponding to the hamiltonian 
\beq 
\hat{H}^\mu= \hat{T}+\hat{W}_{\rm LR}^\mu+\hat{V}^\mu, \qquad \hat{V}^\mu=\sum_i v^\mu(\rv_i).
\label{eq_Hmu}
\eeq
Notice that this multideterminant OEP scheme is different from the one recently
proposed in Ref.~\cite{WeiDelGor-JCP-08}. In Eq.~(\ref{eq_OEPmd}) the weak long-range interaction $\hat{W}_{\rm LR}^\mu$ automatically selects the configuration space needed to yield an accurate solution for the hamiltonian $\hat{H}^\mu$ of Eq.~(\ref{eq_Hmu}), while in Ref.~\cite{WeiDelGor-JCP-08} the configuration space is chosen essentially by hand, using physical and chemical intuition.

In the next Sec.~\ref{sec_mdOverh} we use the Overhauser model to approximate $E^\mu_{\rm c, md}[\rho]$, and we apply our combined formalism to the case of the H$_2$ molecule.
\section{Multideterminant DFT combined with the Overhauser model}
\label{sec_mdOverh}

From the adiabatic connection formalism we can easily write an exact formula for $E^\mu_{\rm c, md}[\rho]$,
\beq
E^\mu_{\rm c, md}[\rho]=\int_\mu^\infty d\mu'\int_0^\infty 4 \pi \,r_{12}^2\, \left[f^{\mu'}(r_{12})-f^{\mu}(r_{12})\right]\frac{2}{\sqrt{\pi}}e^{-\mu'^2 r_{12}^2}\,dr_{12},
\label{eq_ecmd1}
\eeq
which shows that $E^\mu_{\rm c, md}[\rho]$ is determined by the change in the short-range part of the
intracule density when the electron-electron interaction increases from $\erf(\mu r_{12})/r_{12}$ to
the full Coulomb repulsion $1/r_{12}$. By adding and subtracting $f_{\rm KS}(r_{12})$, Eq.~(\ref{eq_ecmd1}) can also be written as
\beq
E^\mu_{\rm c, md}[\rho]=E_c^\mu[\rho]-\int_0^\infty 4\pi\,r_{12}^2\,\left[f^{\mu}(r_{12})-f_{\rm KS}(r_{12})\right]\,\frac{\erfc(\mu r_{12})}{r_{12}}\,dr_{12},
\label{eq_ecmd2}
\eeq
where $E_c^\mu[\rho]$ is the correlation energy of Eq.~(\ref{eq_Eccomp}), defined with respect to the KS determinant.

We computed $f^\mu(r_{12})$  within the Overhauser model, Eq.~(\ref{eq_eff}) with one geminal $g=\sqrt{f}$, using the simple screened potential of Eqs.~(\ref{eq_weff})-(\ref{eq_wc}) with the screening length $\overline{r}_s=R$. For each internuclear distance $R$, the intracules $f^\mu(r_{12})$ have been calculated for 33 values of $\mu$ between $\mu=0.01$ and $\mu=20$. 
By numerical integration we then computed
\beq
\frac{\partial E_c^\mu[\rho]}{\partial \mu}=\int_0^\infty 4 \pi \,r_{12}^2\, \left[f^{\mu}(r_{12})-f_{\rm KS}(r_{12})\right]\frac{2}{\sqrt{\pi}}e^{-\mu'^2 r_{12}^2}\,dr_{12},
\eeq 
and we fitted the values of $\frac{\partial E_c^\mu[\rho]}{\partial \mu}$ with the derivative of the function
\beq
E_c^\mu[\rho]=\frac{a_4}{b^{10}}-\frac{a_1 \mu^6+a_2\mu^7+a_3\mu^8+a_4\mu^{10}}{(1+b^2\mu^2)^5},
\eeq
which has the correct asymptotic behaviors \cite{GorSav-PRA-06}. We also computed, again by numerical integration,  the second term on the right-hand-side of Eq.~(\ref{eq_ecmd2}) in order to obtain $E^\mu_{\rm c, md}[\rho]$.

We then implemented the generalized OEP scheme of Eq.~(\ref{eq_OEPmd}) by first minimizing the effective potential $v^\mu(\rv)$ at the ``generalized-exchange''-only level, and by adding $E^\mu_{\rm c, md}[\rho]$ only as a final correction. Since $E^\mu_{\rm c, md}[\rho]$ is very small, we do not expect substantial changes by implementing a full self-consistent scheme. Our procedure can be summarized with the equation
\beq
E_0=\left(\inf_{v^\mu}\langle\Psi^\mu_{v^\mu}| \hat{T}+\hat{V}_{ee}+\hat{V}_{ne}
|\Psi^\mu_{v^\mu}\rangle\right)+E^\mu_{\rm c, md}[\rho_{\Psi^\mu_{v^\mu}}],
\label{eq_E0md}
\eeq
where $E^\mu_{\rm c, md}[\rho_{\Psi^\mu_{v^\mu}}]$ is calculated with the final density resulting from the minimization in the first term on the right-hand-side of Eq.~(\ref{eq_E0md}).

To carry out the minimization with respect to the potential $v^\mu(\rv)$ in Eq.~(\ref{eq_E0md}) we proceeded as follows. 
We parametrized the potential $v^\mu(\rv)$ with a simple two-parameter form, by adding to the physical external potential a gaussian centered on each atom, $c\, e^{-\gamma r^2}$. The minimization of the expectation $\langle\Psi^\mu_{v^\mu}| \hat{T}+\hat{V}_{ee}+\hat{V}_{ne} |\Psi^\mu_{v^\mu}\rangle$ with respect to the two parameters $c$ and $\gamma$ is done by calculating at each step full-CI wavefunctions $\Psi^\mu_{v^\mu}$ for the hamiltonian with electron-electron interaction $\erf(\mu r_{12})/r_{12}$ and external potential $v^\mu(\rv)$. All calculations were done at the cc-V5Z basis-set level. We also produced with MOLPRO \cite{Molpro-PROG-02} full CI reference results for the physical hamiltonian, for comparison. Our simple parametrization of the  potential $v^\mu$, containing only two parameters, is enough to yield at $\mu=0$ the HF energy within 0.5 mH, which is the accuracy we sought in this study. This way, we avoid all the well-known problems of the OEP method in finite basis set \cite{StaScuDav-JCP-06} at the price of obtaining only an upper bound for our minimization problem (yet, with the reasonable accuracy of 0.5 mH).
\begin{figure}
\includegraphics[width=8.5cm]{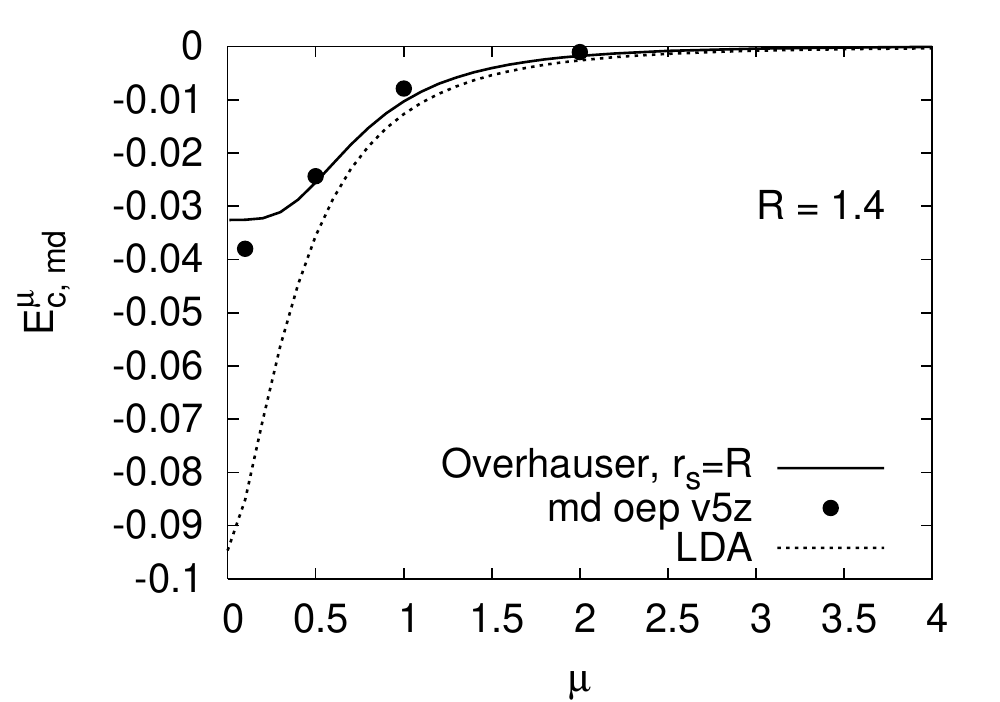}
\includegraphics[width=8.5cm]{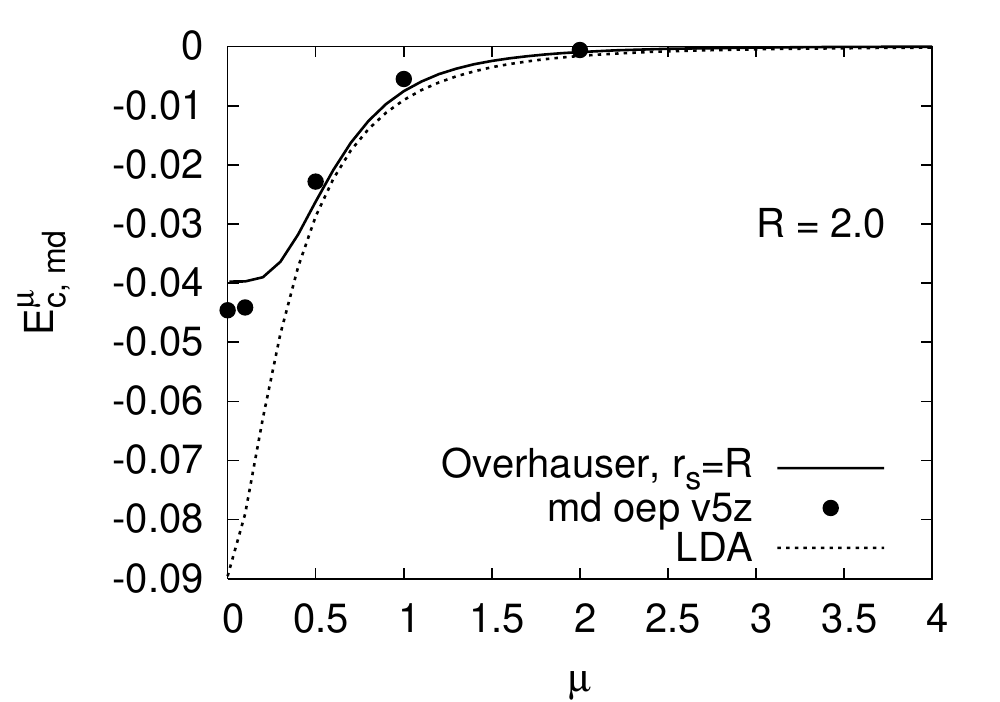}
\includegraphics[width=8.5cm]{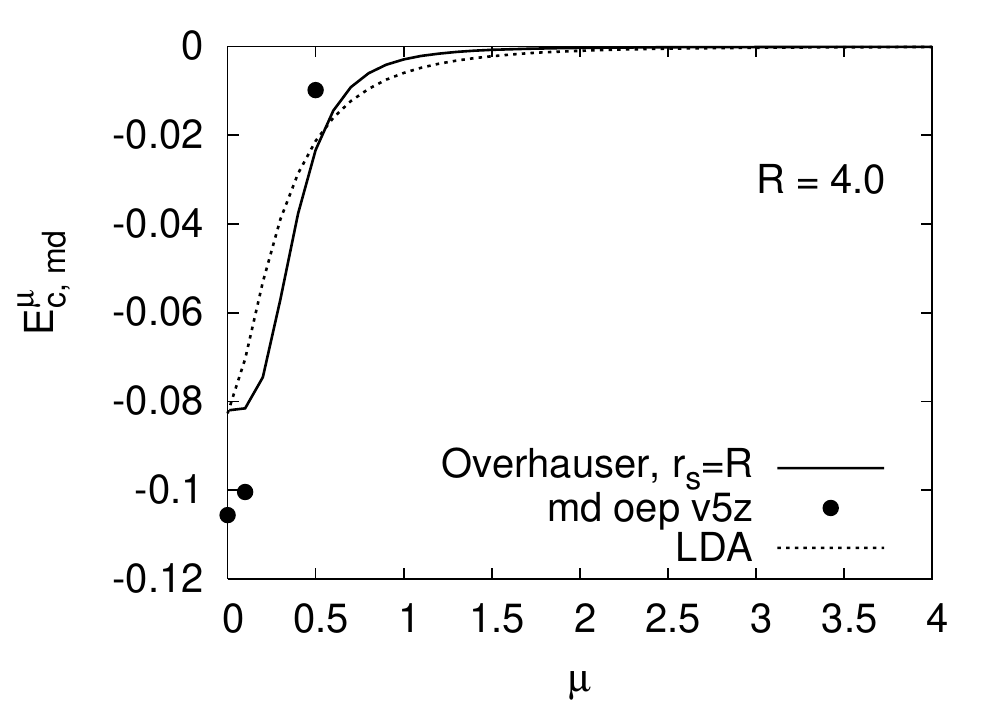}
\caption{The short-range correlation energy for range-separated multideterminant DFT as a function of the cutoff parameter $\mu$ for the H$_2$ molecule at three different values of the internuclear distance $R$. Dots ($\bullet$) are ``exact'' values (see text in Sec.~\ref{sec_mdOverh}), solid lines are the results from the Overhuaser model, and the dashed lines are the LDA values.}
\label{fig_Ecmd}
\end{figure}

In Fig.~\ref{fig_Ecmd} we report the results for $E^\mu_{\rm c, md}[\rho]$ for three different values of the internuclear distance $R$. The dots ($\bullet$) are the ``exact'' values of $E^\mu_{\rm c, md}[\rho]$, i.e., the full-CI total energies obtained with MOLPRO minus the energies corresponding to the first term on the right-hand-side of Eq.~(\ref{eq_E0md}), $\inf_{v^\mu}\langle\Psi^\mu_{v^\mu}| \hat{T}+\hat{V}_{ee}+\hat{V}_{ne}
|\Psi^\mu_{v^\mu}\rangle$. The solid line is $E^\mu_{\rm c, md}[\rho]$ from the Overhauser model, and the dashed line is the LDA result, obtained from the parametrization of Ref.~\cite{PazMorGorBac-PRB-06}, in which $E^\mu_{\rm c, md}[\rho]$ for the uniform electron gas has been calculated with Quantum Monte Carlo  methods. We see from this figure that when the system is still dominated by dynamical correlation, as in the $R=1.4$ and the $R=2$ cases, the Overhauser model yields, even at $\mu=0$ (i.e. for pure KS DFT), correlation energies with errors of $\sim 5$~mH (while LDA is off by $\sim 60$~mH), which reduce to 1~mH at $\mu=0.5$ (where the LDA error is still $\sim 10$~mH).
We focuse here on the value $\mu\sim 0.5$ since it is the one commonly used in practical applications \cite{AngGerSavTou-PRA-05,TouGorSav-TCA-05,GolWerSto-PCCP-05,GolWerStoLeiGorSav-CP-06,GerAngMarKre-JCP-07,FroTouJen-JCP-07,FroJen-PRA-08}.
When the system starts to be dominated by static correlation, as in the $R=4$ case, the Overhauser model with the simple screened potential of Eq.~(\ref{eq_wc}) gives, at $\mu=0$, errors very close to those of LDA ($\sim 20$~mH), which are still of the order of $\sim 10$~mH at $\mu=0.5$. As the molecule approaches the dissociation limit, $R\to\infty$, the exact  $E^\mu_{\rm c, md}[\rho]$ tends to the limiting behavior in which $E^{\mu=0}_{\rm c, md}[\rho]= E_c^{\rm KS}[\rho]$, and $E^{\mu}_{\rm c, md}[\rho]= 0$ for any $\mu>0$. This is due to the fact that, as $R\to\infty$, the long-range only wavefunction $\Psi^\mu$, even at very small $\mu$ (i.e., with an infinitesimal interaction), becomes essentially exact and equal to the Heitler-London wavefunction, so that the functional should be just equal to zero. In this limit, the Overhauser model is wrong for small $\mu$ (because, as explained in Sec.~\ref{sec_Ovh2exact}, it misses the ``overscreening'' at short range), but, for $\mu \gg 0$, yields $E^\mu_{\rm c, md}[\rho]$  that go to zero much faster than LDA, as it can be already grasped from the third panel of Fig.~\ref{fig_Ecmd}. It is thus still more suitable than LDA to be combined with the range-separated multideterminant DFT, but it definitely needs some improvement.

Notice that the Overhauser model would yield much more accurate results if we were able to compute $E^{\mu}_{\rm c, md}[\rho]$ by using in Eq.~(\ref{eq_weff})  instead of $w_{\rm eff}^{\rm KS}(r_{12})$ the interaction $w_{\rm eff}^\mu(r_{12})$ which yields the intracule $f^\mu(r_{12})$ associated to the wavefunction $\Psi^\mu$. This way, we would use the information available in $\Psi^\mu$ to the maximum extent, and we would not have the problems associated to the ``overscreening'' discussed in Sec.~\ref{sec_Ovh2exact}. This possibility will be investigated in future work. 

\section{Conclusions and perspectives}
\label{sec_conc}
We have presented a preliminary study of the combination of range-separated multideterminant DFT with the Overhauser model, with an application to the paradigmatic case of the H$_2$ molecule. We have first analyzed, by means of very accurate variational wavefunctions, the failure of the Overhauser model in describing static correlation and we have then used it to produce an adapted short-range correlation functional for range-separated multideterminant DFT. The results are very good for internuclear distances close to equilibrium, and are still encouraging as the molecule is stretched. Indeed, in the dissociation limit the exact short-range correlation functional should go to zero for any $\mu>0$, and the Overhauser model yields short-range correlation energies that go to  zero faster than LDA as $\mu$ increases. 

Future work will address the study of better approximations for the unknown Overhauser electron-electron interaction, and the development of a more efficient scheme to combine it with range-separated multideterminant DFT. 

\section*{Acknowledgements} 
We thank W. Cencek for providing us with the accurate geminal wave functions of the H$_2$ molecule. 
This work was supported by the ANR (National French Research Agency) under Grant n.~ANR-07-BLAN-0271.

\end{document}